%% file: network_letter_rv_full.tex
\documentclass[%
a4paper,
superscriptaddress,
notitlepage,
bibnotes,
 amsmath,amssymb,
 aps,
10pt,
twocolumn,
 pre,
]{revtex4-2}

 \usepackage{xcolor}
 \colorlet{mylinkcolor}{blue!66!black!80}
  \colorlet{myblue}{black}

\newcommand{\e}{{\rm e}}

\usepackage{mathtools}
 \usepackage[colorlinks=true,linkcolor=mylinkcolor,citecolor=mylinkcolor,filecolor=cyan,urlcolor=mylinkcolor,breaklinks=true]{hyperref}

 \usepackage[utf8]{inputenc}

\newcommand{\avg}[1]{\langle#1\rangle}

\begin{document}
 \title{Violation of Local Detailed Balance Despite a Clear Time-Scale Separation}% Force line breaks with \\
\author{David Hartich}
\email{david.hartich@mpinat.mpg.de}
\affiliation{% 
Mathematical bioPhysics Group, Max Planck Institute for Multidisciplinary Sciences, 37077 Göttingen, Germany}
 \author{Aljaž Godec}%
 \email{agodec@mpinat.mpg.de}
\affiliation{%
Mathematical bioPhysics Group, Max Planck Institute for Multidisciplinary Sciences, 37077 Göttingen, Germany}
\begin{abstract}
   Integrating out fast degrees of freedom is known to yield, to
  a good approximation,  memory-less, i.e.
  Markovian, dynamics. In the presence of such a time-scale separation local detailed balance is believed to
  emerge and to guarantee thermodynamic consistency arbitrarily far
  from equilibrium.  Here we present a transparent example of
a Markov model of a molecular motor where 
  local detailed balance can be violated despite a clear time-scale
  separation and hence 
  Markovian dynamics.
   Driving the system far from equilibrium can lead to
  a violation of local detailed balance against the driving force. We further show that
  local detailed balance can be restored, even in the presence of memory,
if the
coarse-graining is carried out as Milestoning. Our work establishes
Milestoning not only as a kinetically but for
the first time  also as a
thermodynamically consistent coarse-graining method.
Our results are relevant as soon as individual
transition paths are appreciable or can be resolved. 
\end{abstract}
\maketitle

%\emph{Introduction.---}
The formulation of thermodynamic observables,
such as heat and work, along individual stochastic trajectories
unraveled fundamental fluctuation symmetries which matured into the
framework called ``stochastic thermodynamics'' 
\cite{jarz11,seif12,broe15}. In the particular case of continuous-time
 Markov-jump processes the
\textit{local detailed balance} paradigm emerged, relating the kinetics to thermodynamic forces that drive a system out of equilibrium
\cite{katz83, seif11, seif12, broe15}. One inherent assumption of this
paradigm is a separation of timescales \cite{seif11}: the observed
degrees of freedom are slow ensuring that all unobserved/hidden fast degrees of freedom equilibrate with instantaneously
connected (heat or particle) reservoirs
\cite{broe15,yosh21,blom21}.  Accordingly, the forward and corresponding
backward transition rates between a pair of meso-states $A$ and $B$,
$w_{A\to B}$  and  $w_{B\to A}$, respectively,  
are related to the entropy production via \cite{maes21}
\begin{equation}
	k_{\rm B}\ln \frac{w_{A\to B}}{w_{B\to A}}=\text{entropy change $A\to B$},
	\label{eq:LDB}
\end{equation}
where $k_{\rm B}$ is the Boltzmann constant, 
and the entropy difference reflects the change of both, the intrinsic entropy and
the entropy generated in the reservoirs \cite{seif11}.\ However, as soon as slow hidden degrees of freedom emerge (within $A$ or $B$)
 the exact connection between the observed kinetics and the dissipation
 embodied in Eq.~\eqref{eq:LDB} disappears, which was explained
 theoretically \cite{raha07,pigo08,gome08,rold10,pugl10,rold12,espo12,andr12,bo14,dian14,bara14b,zimm15,bo17,kahl18,uhl18,lapo19,lapo20a,lapo21,ehri21,erte22}
 and corroborated experimentally \cite{mehl12}. The equality~\eqref{eq:LDB} can
 nevertheless be restored under specific conditions   
\cite{pugl10,alta12,teza20,hart21}, using affinities \cite{knoc15}, by
stalling the system \cite{pole17,bisk17} or introducing
waiting time distributions \cite{mart19a,skin21,ehri21,skin21a,erte22} that \emph{inter alia} can further trigger anomalous diffusion \cite{hart21a}.

 %both experimentally
 %\cite{mehl12} and theoretically
 %\cite{raha07,pigo08,gome08,rold10,pugl10,rold12,espo12,andr12,bo14,dian14,bara14b,zimm15,bo17,kahl18,uhl18,ehri21},

\begin{figure}
\centering
\includegraphics[width=\columnwidth]{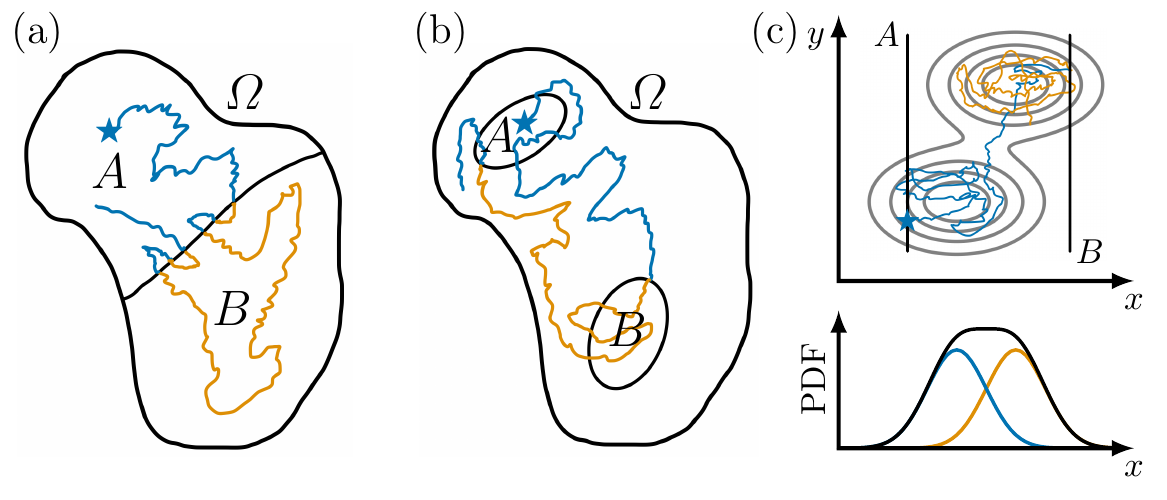}
\caption{Variants of coarse-graining: the color of the full trajectory
  evolving from the blue star represent the instantaneous
  coarse-grained states $A$ (blue) and $B$ (yellow), respectively. (a)
  State lumping: The full set of states $\varOmega$ is decomposed into
  subsets $A$ and $B$. (b) Milestoning based on core sets: two
  metastable states represent the cores $A$ and $B$, and the
  coarse-grained state corresponds to the last visited core. (c) Top: 
%  gray
  contour lines depicting potential iso-surfaces in the $xy$ plane;
  milestones $A$ and $B$ resolve the metastable regions.
  Bottom: measured equilibrium probability
   density function (PDF) with the PDFs of the the individual metastable states indicated in blue and orange, respectively.}	
\label{fig:illustration}
\end{figure}

When the underlying degrees of freedom can assume continuous values
%In the case of continuous underlying coordinates
any coarse-graining
that lumps states as shown in Fig.~\ref{fig:illustration}a
inherently
%fundamentally
%prevents the emergence of
leads to non-Markovian jump dynamics in
continuous time \cite{sari10,schu11} due to fast re-crossings in the
transition region between $A$ and $B$. Notably, these can
nowadays be experimentally resolved
\cite{chun12,chun13,neup12,ritc15,neup16,kim20} and are therefore important 
practically.    
Conversely, \textit{Milestoning} \cite{fara04,shal06} (see
\cite{elbe20,elbe20a,suar21} for a broader perspective) turned out to
be a coarse-graining scheme that allows for a kinetically consistent
mapping of highdimensional dynamics onto a drastically simplified
Markov-jump process \cite{schu11,bere19}. The state space is dissected
into hypersurfaces which may enclose sub-volumes that are called
``cores'' \cite{schu11,bere19}.  Fig.~\ref{fig:illustration}b depicts
two such cores $A$ and $B$, whereby the color of the trajectory encodes the
last visited core. Beyond a short transient, Markov-jump dynamics emerges from the
coarse-graining whenever the trajectory upon leaving any core either
(i) quickly returns to it or (ii) quickly transits to the next core
\cite{schu11}. Hereby, condition (i) ensures a local equilibration
prior to leaving a state that is required for the emergence of local
detailed balance \cite{katz83, seif11, seif12, broe15}. 
Besides being kinetically consistent,
Milestoning offers two main advantages over lumping.
 
First, in experiments probing
low-dimensional observables one may be able to separate pairs of
metastable states even if their projections onto the observable
%display an
%(partially)
overlap \cite{nage19}.\ This is illustrated in
  Fig.~\ref{fig:illustration}c, where two
  seemingly overlapping metastable states in the projected space $x$ are
  resolved by choosing the respective milestones outside the overlapping
  region.\ Whenever a milestone is left, the trajectory rapidly returns
%(i.e., does not move to the other metastable state)
or quickly transits to the other milestone.\ Thus, the last visited
milestone to a good approximation reflects the currently visited
metastable region in a possibly higher-dimensional (here 2d)
underlying space.\ 
Second,   
we recently discovered that Milestoning naturally ensures local
detailed balance in the presence of a time-scale separation
\cite{hart21}.\ Surprisingly, this extends even to systems without a
clear time-scale separation, which we investigate further below.\
Notably, with so-called ``dynamical coring''  \cite{jain14,nage19} one
can, under certain conditions, convert a ``lumped''
process into a ``milestoned'' process by manually
discarding short recrossing events as those shown in
Fig.~\ref{fig:illustration}a.

In contrast to continuous-space processes, the lumping of dynamics that
evolve on a
discrete state space \cite{seif12,broe15,katz83,maes21,raha07,pigo08,rold10,pugl10,rold12,espo12,andr12,bo14,dian14,bara14b,zimm15,bo17,kahl18,uhl18,ehri21,alta12,teza20,mart19a,skin21,skin21a,pole17,bisk17,blom21,yosh21,erte22} %(kahl18)
\textit{can} in fact yield an effectively %continuous time
Markovian jump process.\ According to perturbation theory %was used to show that
Eq.~\eqref{eq:LDB} is satisfied by lumped-state dynamics in the
limit of an infinite time-scale separation \cite{espo12}, which was
corroborated in \cite{dian14,bara14b,bo17}.  
This general belief was, however, never systematically scrutinized in practice.

 In this Letter we show, by means of a simple yet biophysically
 relevant example, that time-scale separation surprisingly and against
 common belief does
 \textit{not} %suffice for the emergence %!<- sonts haben wir zu oft "suffice"
ensure the existence
 of local detailed balance.
 The minimum time-scale separation required for Eq.~\eqref{eq:LDB} to
 hold may grow exponentially with the thermodynamic driving force. In
 other words, time-scale separation may not suffice
 \textit{arbitrarily far} from equilibrium. 
 Milestoning, in stark contrast to lumping (see
 Fig.~\ref{fig:illustration}), robustly ensures local
 detailed balance in the limit of a time-scale separation. This result
 indicates that unlike lumping, Milestoning generically yields a thermodynamically
 consistent coarse-graining.

\begin{figure}
\centering
\includegraphics[width=\columnwidth]{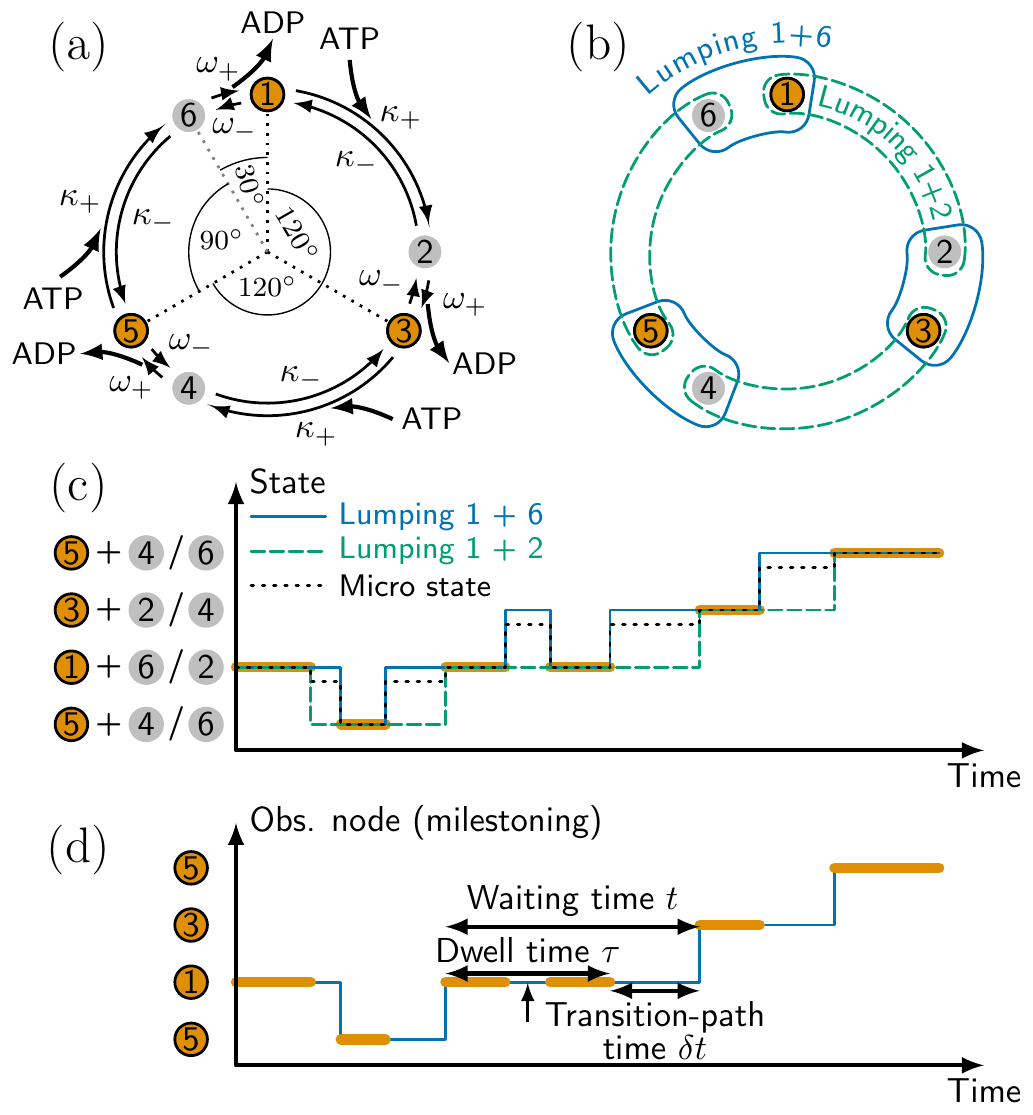}
\caption{Model and coarse-graining. (a)~Full six state model; the
  dotted lines denote odd rotational states $\{1,3,5\}$ (orange) separated by
  $120^\circ$. The even intermediate states (gray) separate each
  rotation step into $90^\circ$ and $30^\circ$ substeps. (b)~Two types of lumping; the solid boxes ``lumping 1+6'' lump states $\{1,6\}$, $\{2,3\}$, and
$\{5,6\}$, respectively, whereas the dashed boxes ``lumping 1+2'' lump
  states $\{1,2\}$, $\{3,4\}$, and $\{5,6\}$. (c)~Coarse-grained
  trajectory using ``lumping 1+2'' and ``lumping 1+6''; the orange
  %thick
  segments represent visits of odd states. The dotted line (``micro state'') indicates the rotational state of the motor as function of time. (d)~Coarse-grained
  trajectory deduced from Milestoning; the milestones are placed
  %at angles that correspond
  at odd states.}
\label{fig:schematics}
\end{figure}

\emph{$F_1$-ATPase driven far from equilibrium.---}We consider the molecular motor F${}_1$-ATPase %whose rotation is
driven by the hydrolysis of adenosine triphosphate (ATP). The dynamics
evolves as a Markov processes on six rotational states  \cite{yasu01}
as shown in Fig.~\ref{fig:schematics}a: 
The binding
of ATP occurs with a rate $\kappa_+$ proportional to the
concentration of ATP and effetcs a $90^\circ$ rotation. The reverse
unbinding occurs with the rate $\kappa_-$. ATP hydrolisis to ADP is assumed to be infinitely fast. The release of ADP
%is assumed to
occurs with  %a finite
rate $\omega_+$ and triggers a $30^\circ$ rotation, and the reverse step
occurs with rate $\omega_-$.

The free energy $\mu$ liberated by the
hydrolysis of one $\text{ATP}\to \text{ADP}$ at a
given concentration relates to the entropy
change times the
 temperature $T$, and local detailed balance \eqref{eq:LDB} imposes
\begin{equation}
	k_{\rm B}T\ln \frac{\omega_+\kappa_+}{\omega_-\kappa_-}=\mu.
	\label{eq:mu}
\end{equation}
Heneceforth we measure energies, $\mu$, in units of the thermal energy $k_{\rm B}T$.
 The steady state probability to find the ATPase in even and odd states is given by \cite{schn76}
\begin{equation}
	P_{\rm odd}=\frac{\omega_++\kappa_-}{\omega+\kappa} \quad\text{and}\quad P_{\rm even}=\frac{\omega_-+\kappa_+}{\omega+\kappa},
\end{equation}
respectively,
where we defined $\kappa\equiv\kappa_++\kappa_-$ and $\omega\equiv\omega_++\omega_-$. The entropy production rate
can be expressed with the rate of ATP consumption,
$J= P_{\rm odd}\kappa_+-P_{\rm even}\kappa_-$, via  \cite{schn76}
\begin{equation}
	\sigma=J\mu.
	\label{eq:EP}
\end{equation}
This completes the description of the ``full'' system.

\emph{Lumping.---}We now perform a coarse-graining to reduce the six states to
three. Two sensible ways to lump the states are shown in Fig.~\ref{fig:schematics}b.
Assuming Markovian dynamics the effective forward ``$+$'' and backward ``$-$'' rates on the lumped
space read \cite{espo12}
\begin{equation}
\begin{aligned}
	 W^{1+6}_\pm&= P_{\substack{\text{odd}\\\text{even}}}\kappa_\pm=\frac{(\omega_\pm+\kappa_\mp)\kappa_\pm}{\kappa+\omega},\\
  W^{1+2}_\pm&= P_{\substack{\text{even}\\\text{odd}}}\omega_\pm
  =
  \frac{(\omega_\mp+\kappa_\pm)\omega_\pm}{\kappa+\omega},
\end{aligned}
\label{eq:cg_rates}
\end{equation}
and satisfy $J= W^{1+6}_+-W^{1+6}_-=W^{1+2}_+-W^{1+2}_-$.
In terms of effective rates the coarse-grained entropy 
reads \cite{espo12}
\begin{equation}
	\tilde \sigma_z=J\ln \frac{W^z_+}{W^z_-},
	\label{eq:EPcg}
\end{equation}
with $z=1+6$ or $z=1+2$ and 
using Eqs.~(\ref{eq:EP}-\ref{eq:EPcg})
yields
\begin{align}
	\frac{\tilde\sigma_{1+6}}{\sigma}
%	=1-\frac{1}{\mu}\ln\frac{1+\kappa_+/\omega_-}{1+\kappa_-/\omega_+}
&=1-\frac{1}{\mu}\ln\frac{1+\e^\mu\kappa_-/\omega_+}{1+\kappa_-/\omega_+},
\label{eq:s16}\\
%\end{equation}
%and 
%\begin{equation}
	\frac{\tilde\sigma_{1+2}}{\sigma}
%	=1-\frac{1}{\mu}\ln\frac{1+\omega_+/\kappa-}{1+\omega_-/\kappa_+}
&=1-\frac{1}{\mu}\ln\frac{1+\e^\mu\omega_-/\kappa_+}{1+\omega_-/\kappa_+}.
\label{eq:s12}
\end{align}
Both ratios \eqref{eq:s16} and
\eqref{eq:s12} are positive and bounded by $1$ \cite{espo12}, i.e.,
$\tilde\sigma_z\le\sigma$ (see also \cite{nguy17}). 

\emph{Time-scale separation.---}In agreement with \cite{espo12} (see
also \cite{dian14,bara14b,bo17}) in the limit 
$\kappa_-\ll \e^{-\mu}\omega_+$ (i.e.,
$\kappa\to 0$) we
obtain $\tilde \sigma_{1+6}\approx \sigma$, whereas  the limit $\omega_-\ll \e^{-\mu}\kappa_+$
(i.e., $\omega\to 0$) yields $\tilde \sigma_{1+2}\approx \sigma$. In
other words, when %the
hidden jumps are much faster than those between
lumped states, the coarse-grained dynamics are approximately Markovian
\emph{and} preserve the entropy production.

\textcolor{myblue}{A time-scale separation is manifested as a gap
  in the spectrum of the Markov generator, which separates fast from
  slow modes (see Part~II in \cite{yin98}).} In our model $\kappa\gg \omega$ and $\kappa\ll\omega$ are the only kinds of time-scale separation,
and in principle require two different types of lumping
\textcolor{myblue}{(for details see \footnote{The Supplemental Material
  provides a detailed discussion of the time-scale separation and provides additional examples which cite Refs.~[69--81].})}.  
At high ATP concentration ($\kappa\gg \omega$) ``lumping $1+2$'' (see
dashed boxes in Fig.~\ref{fig:schematics}b) hides the fast degrees
freedom $\sim\kappa$.  Conversely, at low ATP concentration one should
rather lump $1+6$ (see solid boxes in Fig.~\ref{fig:schematics}b). 
Note that whenever the entropy production rate is deduced from a
master equation
\cite{seif12,broe15,katz83,maes21,raha07,pigo08,rold10,pugl10,rold12,espo12,andr12,bo14,dian14,bara14b,zimm15,bo17,kahl18,uhl18,ehri21,alta12,teza20,mart19a,skin21,skin21a,pole17,bisk17,blom21,yosh21,erte22}
one explicitly (or implicitly) assumes the observed degrees of
freedom to be formally infinitely slower than any possibly hidden ones. 

\begin{figure}
\centering
\includegraphics[width=\columnwidth]{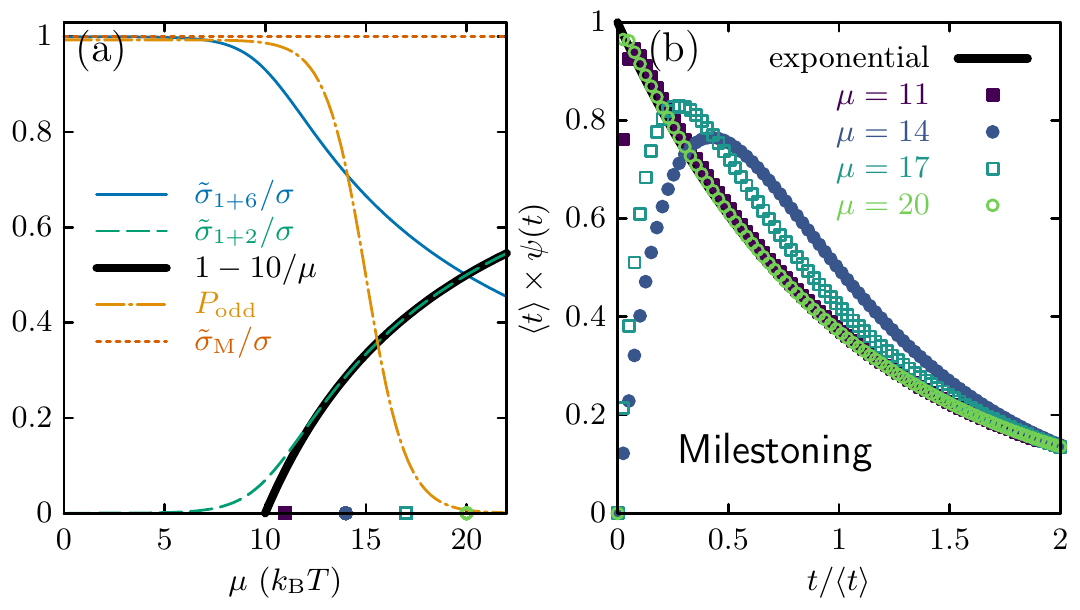}
\caption{Entropy production and waiting time statistics with
  Milestoning. (a) Entropy production, $\tilde\sigma_{1+6}$ and
  $\tilde\sigma_{1+2}$  deduced from the lumpings ``1+6'' and ``1+2'',
  respectively.  The thick line depicting $1-10/\mu$ approaches
  $\tilde\sigma_{1+2}/\sigma$ in the limit $\mu\to\infty$. The dash-dotted
  yellow line denotes the steady state probability $P_{\rm
    odd}$. (b)~Probability density of waiting time $t$ (see
  Fig.~\ref{fig:schematics}d) for $\mu$ values indicated in (a). The
  thick black line depicts an exponential density $\propto \e^{-t/\avg{t}}$.
   Parameters: $\omega_+=1$, $\omega_-=\e^{-5}$, $\kappa_+=\e^{\mu-15}$, and $\kappa_-=\e^{-10}$.}
\label{fig:entropy}
\end{figure}

\emph{Violation of local detailed balance.---}In practice
an infinite time-scale does not exist and the driving $\mu$ becomes
important if it substantially exceeds the thermal energy ($\mu\gg1$),
which in turn implies $\e^\mu\ggg1$.  To see this set $\omega_\pm$ and
$\kappa_-$ to be constant while varying the ATP concentration as
$\kappa_+\propto e^{\mu}$ as in \cite{yasu01} (the parameters are
given in Fig.~\ref{fig:entropy}).  For $\mu<10$ we find $\omega\gg
\kappa$ and as expected $\tilde \sigma_{1+6}\approx \sigma$ (see
Fig.~\ref{fig:entropy}).  For $\mu>15$ we have $\omega\ll \kappa$,
however, to our surprise $\tilde \sigma_{1+2}\not\approx \sigma$
(because $\omega_-\not\ll \e^{-\mu}\kappa_+$). 
Inspecting Eq.~\eqref{eq:s16} we actually find $\tilde\sigma_{1+2}\approx 1-10/\mu$ (see Fig.~\ref{fig:entropy}a).
Thus one obtains $\tilde \sigma_{1+2}/\sigma\to 1$ in the limit
$\mu\to \infty$, which is approached algebraically slowly. For example,
in the already unphysical situation
$\mu=40\,k_{\rm B}T$  \footnote{At physiological conditions
$\mu\approx 20$ the ATP concentration is $c\sim 1\,$mMol. Thus,
$\mu=40$ clearly corresponds to an unphysical ATP concentration of
$c\sim0.001\e^{20}\approx 500000\,$Mol.} only $75\,\%$ of the entropy production are recovered in Eq.~\eqref{eq:EPcg}.
Moreover, at physiological conditions $\mu=20$ we find a clear
time-scale separation, $\kappa/\omega\approx140\gg1$ [see $\lambda_1$ and $\lambda_2$ in Eq.~\eqref{eq:numeric} for the precise time-scales],, yet the
entropy production is not even remotely restored. This surprising
finding is the first main result of this Letter. 

How can we reconcile this?  For convenience we focus on $\mu=20$ and
the lumping ``1+2'', which in fact represents a semi-Markov process of second order \cite{mart19a} (see also \cite{skin21,ehri21,skin21a}). That is,
the waiting time density $\psi_{\pm|\pm}(t)$ depends on both, the previous and
next visited state
%direction $i=\pm$ from where the lumped state was entered and the
%direction $j=\pm$ where it is heading
with the normalization $\int_{0}^\infty [\psi_{+|i}(t)+\psi_{-|i}(t)]
dt=1$ and $i=\pm$. In particular, for the given parameters we find
\begin{align}
	\hspace{-3mm}\begin{pmatrix}
		\psi_{+|+}(t)\!&\psi_{+|-}(t)\\
		\psi_{-|+}(t)\!&\psi_{-|-}(t)
	\end{pmatrix}
	&\!\approx\!\begin{pmatrix}
		1.007&1.000\\
		2.089\cdot 10^{-9}&2.075\cdot 10^{-9}
	\end{pmatrix}\e^{-\lambda_1t}\nonumber\\
	+&
	\begin{pmatrix}
		-1.007 & 3.100\cdot 10^{-7}\\
		6.738\cdot 10^{-3}& -2.075\cdot 10^{-9}
	\end{pmatrix}\e^{-\lambda_2t},
	\label{eq:numeric}
\end{align}
where
$\lambda_1\approx1.0\textcolor{myblue}{{}\approx\omega}$ and $\lambda_2\approx 148.4\textcolor{myblue}{{}\approx\kappa}$.
The analytical expression for the waiting time density is
immaterial for the present
discussion but straightforward to determine. 
For times $t\gtrsim 0.15$ the jumps are essentially Markovian --
the waiting time density is to a good
approximation exponential and independent of the previous step, $\psi_{\pm|+}\approx \psi_{\pm|-}$, and the fast decaying mode is
negligible, $\e^{-\lambda_2t}\approx 0$.
\textcolor{myblue}{Remarkably, using Eq.~\eqref{eq:numeric} one finds $\ln[\psi_{+|i}(t)/\psi_{-|i}(t)]\approx \mu=20$ for  for all $t\gtrsim 0.15$ and $i=\pm$.}
Hence, only short times $t\le 0.15$ encode a violation of
Markovianity \textcolor{myblue}{\emph{and} broken local detailed balance}.\ At strong driving most of the jumps occur in positive
direction ``$+$'' and on average take equally long
$\approx 1/\lambda_1\approx1$.
In fact, only the backward jump ``$-$'' can be faster
on average, however, if and only if the preceding jump occurred in the forward
``$+$'' direction, i.e. a forward transition is followed immediately
by a backward transition. In this case one finds $\int_{0}^\infty
t\psi_{-|+}(t)d t/\int_{0}^\infty \psi_{-|+}(t)d
t\approx0.0067\approx1/\lambda_2$. These rare events lead to an
``overestimation''  of the effective backward transition rate
$W_-^{1+2}\gtrsim \omega_-\kappa_+/\kappa $. Note that a locally
equilibrated backward rate would need to satisfy
$\ln(W_-^{1+2})\approx\ln(\omega_-\kappa_+/\kappa)$\textcolor{myblue}{,
  which is satisfied if in addition to the time-scale separation the individual rates satisfy $\kappa_\pm\gg\omega_\pm$ (see, e.g., \cite{espo12})}.
  
By evaluating exactly the waiting time distribution to include the short-time 
behavior one is able to restore the entropy production from the two-step affinity via \cite{mart19a}
  \begin{equation}
\sigma_{\rm aff}^{1+2}=J\ln\frac{\int_0^\infty\psi_{+|+}(t)d t}{\int_0^\infty\psi_{-|-}(t)d t}=J\mu,
\label{eq:sigma_aff}
\end{equation}
where the last equality follows from Eqs.~\eqref{eq:mu} and \eqref{eq:numeric} (here $\mu=20$). 
Thus, by taking into account the tiny non-Markovian features in
Eq.~\eqref{eq:numeric} one can in principle recover the entropy
production. This, however, poses a serious practical problem at strong driving
$\mu\gg1$. Namely, to deduce Eq.~\eqref{eq:sigma_aff} from an experiment we formally
require a trajectory with statistically sufficiently many incidents of finding \emph{two consecutive backward steps not interrupted
by a forward step}. It thus seems that one is required to reliably
observe rare events with a probability $\propto(\e^{-\mu})^2$, which
may not be feasible.

 In the following we illustrate how an alternative coarse-graining --
 Milestoning -- effectively restores Markovian dynamics in a thermodynamically consistent manner
 while it concurrently effectively squares the sample size by relying
 only on the evaluation of single rare backward jumps that occur with
 probability $\approx \e^{-\mu}$.

\emph{Thermodynamic consistency of Milestoning.---}We define
three milestones (or cores) at locations highlighted by dotted black
lines in Fig.~\ref{fig:illustration}a. These represent the three odd
rotational states. We measure the passages across the milestones
(see thick yellow lines in Fig.~\ref{fig:illustration}c). If the angle
were measured continuously, the passages through the milestones would
correspond to instantaneous events \cite{hart21}. The coarse-grained
process at any time reflects the last visited Milestone (see blue
line). As in Ref.~\cite{hart21} we dissect waiting times into the
dwell and transition time periods. The dwell time represents all loops
returning to the original milestone, while the transition-path time
reflects the time of commuting between milestones. The waiting time
can be shown to be the sum of the statistically independent dwell and
transition-path times (see second main result in \cite{hart21}). The
main advantage of this decomposition is that the statistics of
transition-path time encode information about potentially hidden
multidimensional pathways \cite{sati20} (see also
\cite{maka21,hart21,bere21}). 

If the gaps between revisitations of the same milestone (see vertical
arrow in Fig.~\ref{fig:illustration}d) \emph{and} transition-path
times are negligibly short compared to the waiting time in a state,
the resulting ``Milestoned process'' becomes, to a good approximation,
Markovian \cite{schu11}.  Note that milestones may represent
closed (see \cite{schu11} and
Fig.~\ref{fig:illustration}b) or open (see \cite{fara04,shal06} and
Fig.~\ref{fig:illustration}c) hypersurfaces.

Let  $\phi_\pm$ denote the splitting probability that the next
milestone will be visited in the forward ``$+$'' and backward ``$-$''
direction, respectively. One can confirm (cf. first main result in
\cite{hart21}) that 
\begin{equation}
	\ln\frac{\phi_+}{\phi_-}=\ln\frac{\kappa_+\omega_-}{\kappa_-\omega_-}=\mu
	\label{eq:aff_M}
\end{equation}
holds.  That is, Milestoning transition probabilities \emph{exactly}
encode the entropy production per hydrolyzed ATP.   

Since transition-path times obey a reflection symmetry \cite{bere06}
and because the dwell time statistics do not depend on the exit
direction \cite{hart21} the waiting time densities in the $+$ and $-$
direction coincide, i.e. $\psi_\pm(t)=\psi(t)$. In the presence of
hidden dissipative mechanisms the symmetry may be lifted counterintuitively \cite{glad19,ryab19}. 
 Denoting the mean waiting time by $\avg{t}=\int_0^\infty t\psi(t)dt$,
 the steady state current becomes $J^{\rm M}=\phi_+/\avg{t}-\phi_-/\avg{t}=J$.
 Defining the Milestoning rates as $W^{\rm M}_\pm=\phi_+/\avg{t}$ and
 inserting them into Eq.~\eqref{eq:EPcg} yields, using Eqs.~\eqref{eq:EP} and
 \eqref{eq:aff_M}, $\tilde\sigma_{\rm
   M}=\sigma$. Thus, Milestoning in contrast to lumping
 \emph{preserves} the entropy production in the limit of
 a time-scale separation \emph{and} beyond.  
 
 Upon inspecting the waiting time density we find that it is to a good
 approximation memory-less for $\mu\lesssim 10$ as well as for
 $\mu\gtrsim 20$, while the non-exponential behavior is most pronounced
 in the regime  $10\le \mu\le 20$ (see Fig.~\ref{fig:entropy}b).
 Thus, in the limit of either of the two time-scale separations, $\mu\lesssim 10$
 and $\mu\gtrsim 20$, the Milestoned dynamics is to a good
 approximation Markovian. In contrast to lumping, Milestoning restores
 local detailed balance \eqref{eq:LDB} in both directions, parallel and anti-parallel to the driving, even at large asymmetries, which is the second main result of this Letter.

 Notably, the regime $\mu\lesssim 10$ clearly fulfills both criteria
 (i) and (ii) for the
 emergence of Markovian dynamics \cite{schu11}
 %(see also second paragraph in the Introduction)
 if the probability to reside within a
 core satisfies $P_{\rm odd}\approx 1$. Conversely, the opposite limit
 $\mu\gtrsim 20$ does \emph{not} obviously imply Markovian
 kinetics. To understand why it does so nevertheless, we point out that  
 in this limit (a)
  $P_{\rm even}=1-P_{\rm odd}\approx 1$. 
 If we were to choose the even (gray) states as cores instead of the odd
 (yellow) ones (see Fig.~\ref{fig:schematics}a),
 we would obviously restore the criteria for the emergence of
 Markovian dynamics \cite{schu11}. It turns out further that
 (b) the waiting 
 time density remains unaffected by the exchange of $\omega_\pm$ and
 $\kappa_\pm$, i.e. it does not depend on whether we choose the odd
 or even states as milestones. This explains why an exponential
 distribution emerges to a good approximation also in the limit
 $\mu\gtrsim20$. We also note that the kinetic hysteresis discovered in \cite{hart21}
 almost vanishes as soon as Markovian dynamics emerge
 \emph{and} the aforementioned criteria \cite{schu11}
 are satisfied, which here follows from (a) by choosing the even states as milestones.

\emph{Conclusion.---}We have shown that a clear time-scale
separation, in contrast to the common belief, is only a necessary but
\emph{not} a sufficient condition for the validity of local detailed
balance.
%We explicitly showed that the lumping of fast states leads to
%essentially Markovian dynamics.  <- Wird auch noch präziser wiederholt
By %means of an example of
coarse-graining a detailed Markov model of a strongly driven molecular motor we
demonstrated %the existence of
a clear time-scale separation between the observed and hidden degrees of
freedom and hence Markovian dynamics of the observable, and concurrently
the non-existence of a local equilibrium against the driving.
%Milestoning, in contrast to lumping, was shown to restore local detailed
%balance in the presence of a time-scale separation and beyond. %<-
%Der Satzt ist hier praktisch 1:1 wiederholt daher raus
Our work demonstrates, for the first time, that Milestoning restores
thermodynamic consistency in the steady state in the presence of strong driving even if
the dynamics displays memory.
A coarse-graining based on lumping may yield effectively
Markovian dynamics that nevertheless violates local detailed
balance. It will be interesting to revisit 
  recent works on the thermodynamics of systems with slow hidden
  degrees of freedom that employed lumping
\cite{mart19a,skin21,ehri21,skin21a} to inspect if and how these
  change under the thermodynamically consistent
  Milestoning which will lead to correlated transitions
  \cite{hawk11} and/or dwell times
  \cite{hart21}. \textcolor{myblue}{Beyond the examples shown here as
    well as
    in Sec.~II of \cite{Note1} it will be interesting to investigate
    whether the two conditions for Markovianity together with Milestoning \cite{schu11} generally guarantee the validity of local detailed balance.}

 % \begin{acknowledgments}
\textbf{Acknowledgments.}
 The financial support from the German Research Foundation (DFG) through the Emmy Noether Program GO 2762/1-2 to A. G. is gratefully acknowledged.

% \appendix
 \section*{Supplemental Material}
 \section{Time-scale separation in the ATPase model}
 \subsection{General discussion}
  The time-scale separation emerges if any of the two clusterings
  ``Lumping 1+2'' or ``Lumping 1+6'' have meso-states that relax much
  faster than the remaining transitions. %corresponding other one.
  Since two-level systems relax to equilibrium with a rate that is
  roughly given by the sum of both rates connecting the two states, we
  obtain a time-scale separation as soon as either $\omega\ll\kappa$
  or $\omega\gg\kappa$, which we explain more thoroughly in the
  following. 
   
Without loss of generality we focus on ``Lumping 1+2'' (see Fig. 2b in
the main text) and rationalize why $\kappa\gg \omega$ actually
corresponds to a proper separation of time scales.
 The other time-scale separation associated with ``Lumping 1+6'' and
 $\omega\gg \kappa$ follows by analogy, which is mathematically
 obtained by interchanging $\omega_\pm\leftrightarrow\kappa_\pm$ in
 the discussion below.

 The dynamics inside a mesostate $1+2$ (see Fig. 2b in the main text)
 occurs with ``internal'' rates $\kappa_\pm$ within the
 lumped-states. The lumped states are exited either with rate
 $\omega_-$ from state $1$ (odd numbered) or with rate $\omega_+$ from
 state $2$ (even numbered). 
 The eigenvalues of the Generator within one lumped state represent the zeros of the characteristic polynomial
 \begin{equation}
 \chi(\lambda)=	\det
 	\begin{pmatrix}
 		\lambda +\kappa_++\omega_-&-\kappa_-\\
 		-\kappa_+&\lambda +\kappa_-+\omega_+
 	\end{pmatrix}
 \end{equation}
 which leads to the eigenvalues
 \begin{equation}
 	\lambda_\pm=\frac{\omega+\kappa}{2}
 	\pm\sqrt{\Big(\frac{\omega+\kappa}{2}\Big)^2-C},
 \end{equation}
 where $\omega\equiv\omega_++\omega_-$, $\kappa\equiv\kappa_++\kappa_-$ and $C\equiv \omega_+\kappa_++\omega_-\kappa_-+\omega_+\omega_-$. In the main text we set $\lambda_2=\lambda_+$ and $\lambda_1=\lambda_-$.
 The two relaxation time scales are $1/\lambda_+$ and $1/\lambda_-$,
 which correspond to the two-scale exponential decay due to the two states
 within one meso-state. In general, lumping $n$ states will lead to
 $n$ exponentially decaying modes. 
 A separation of time scale demands that all dynamical modes except
 for one exponentially decaying mode are quickly relaxing (here $\lambda_+\gg\lambda_-$ or $1\gg \lambda_-/\lambda_+$). We relate this condition to the rates via
 \begin{align}
 1&\stackrel{!}{\gg}\frac{\lambda_-}{\lambda_+}
= \frac{\lambda_-\lambda_+}{\lambda_+^2}	
\simeq \frac{\lambda_-\lambda_+}{(\lambda_++\lambda_-)^2}\nonumber\\
&=\frac{\omega_+\kappa_++\omega_-\kappa_-+\omega_+\omega_-}{(\omega+\kappa)^2}\equiv R,
	\label{eq:R_def}
 \end{align}
where in the last step of the first line we self-consistently re-used
the first inequality implying $\lambda_+\simeq\lambda_++\lambda_-=\omega+\kappa$, and the very last step defines the ratio $R$. In other words, time-scale separation for ``Lumping 1+2'' demands  $R\ll1$. Let us now test whether (and when) this condition is satisfied in our model.

 \subsection{Testing the time-scale separation in the model}

Defining $p\equiv\kappa_-/\kappa$ and $q\equiv \omega_-/\omega$, which both satisfy $0\le p,q\le 1$,
the ratio \eqref{eq:R_def} becomes
\begin{equation}
	R=\frac{[(1-p)(1-q)+pq]\omega\kappa+q(1-q)\omega^2}{(\omega+\kappa)^2}.
	\label{eq:R}
\end{equation}
If $\kappa\gg \omega$ we find
$R\approx [(1-p)(1-q)+pq]\omega\kappa^{-1}\ll 1$, which using
$q\approx0.0067\ll 1$ from the parameter given in Fig.~3 in our case
yields $R\approx (1-p)\omega/\kappa\le  \omega/\kappa$. Thus $\kappa
\gg \omega$ clearly implies a separation of time scales, i.e., $R\ll
1$,  which is further ``strengthened'' due to $(1-p)<1$. In other
words, $\kappa\gg \omega$ implies that all internal modes relax much
faster than the slowest one, which reflects \emph{a memoryless
exponential decay}. This completes the proof that $\kappa\gg \omega$
in fact implies a separation of time scales $\lambda_+\gg\lambda_-$
for Lumping $1+2$. Note that we have determined in Eq.~(9) in the main
text the time-scale separation
$R=\lambda_1/\lambda_2\equiv\lambda_+/\lambda_-\simeq 1/148$ that is actually stronger than $\omega/\kappa\simeq1/140$.

\subsection{Spectral criterion for time-scale separation}

\begin{figure}
	\centering
	\includegraphics{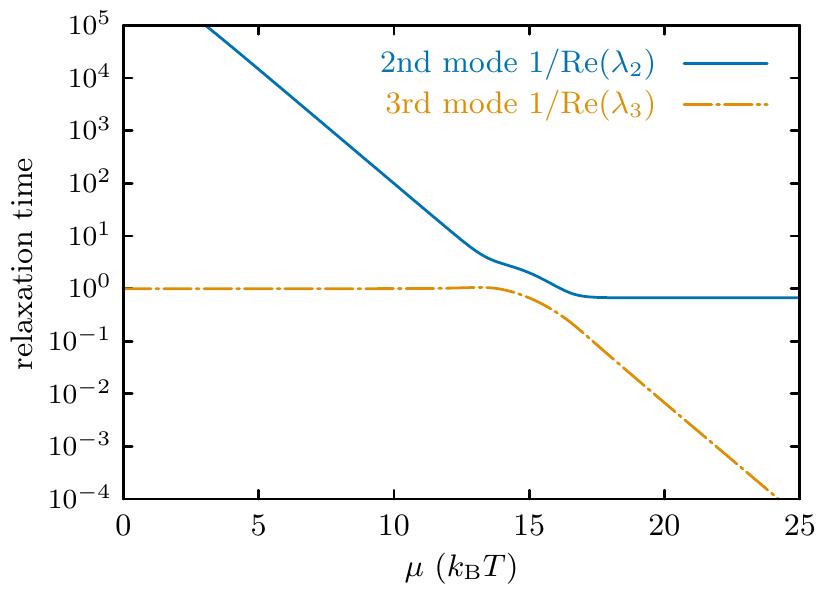}
	\caption{Relaxation time of second and third eigenmode as function of the driving $\mu$. }
	\label{fig:gap}
\end{figure}

For the sake of completeness we discuss the time-scale separation also
in a slightly alternative (but equivalent) interpretation using the
spectrum of the Markov generator \cite{moro95,yin98} (see also
Ref.~\cite{fala21} for recent related work on local detailed balance). In this setting we consider the full system (in our case the six-state system from Fig.~2(a) in the main text). 
For simplicity we focus on the specific example with parameters given
in Fig. 3 and $\mu=20$, which is also used in Eq. (9) in the main
text.  Inserting all the rates in the \emph{full}
six-state system leads to to six eigenvalues, where $\lambda_0=0$ corresponds to the steady state distribution that does not change with time. The other five eigenvalues are $\lambda_{1,2}\approx 1.5\pm 0.88\,{\rm i}, \lambda_{3,4}\approx147.9\pm 0.88\,{\rm i}$, $\lambda_5=149.4$.
Consistent with the 
 previous discussion these values confirm the gap in the eigenvalue spectrum
 $0=\mathrm{Re}(\lambda_0)\le \mathrm{Re}(\lambda_1)\le  \mathrm{Re}(\lambda_2)\ll \mathrm{Re}(\lambda_3)\le \mathrm{Re}(\lambda_4)\le  \mathrm{Re}(\lambda_5)$,
 i.e., three eigenmodes ($0^{\rm th}$-$2^{\rm nd}$ eigenvalue) relax much slower with a rate $\le 1.5$ than the other three (3rd-5th eigenvalue) with a rate $\ge149.4$. We thus confirm the time-scale separation by means of the definition given in
Ref.~\cite{moro95}. More generally, we explicitly illustrate  in Fig.~\ref{fig:gap} the size of the gap between the second and third mode as function of the driving $\mu$, which except for $\mu\sim 15$ is in fact always quite pronounced.
 
 Note that a (coarse-grained) three-state Markov model can account
 only of three eigenvalues $\lambda_0,\lambda_1,\lambda_2$ (and hence
 time scales) and their corresponding eigenfunctions. If more time
 scales are required/desired one must increase the number of states.

\subsection{Pecularity of the violation of local detailed balance}
In Eq.~(9) in the main text we evaluate the waiting time in the
coarse-grained state given that the next move and preceding moves are
heading in either of both directions ($+$ or $-$), respectively.
Inspecting said waiting time density one immediately finds that local
detailed balance is satisfied, to a very good approximation, if the
second term is ignored. Thus, the second term alone encodes the
violation of local detailed balance. 

Since the second term decays much faster than the first term ($\lambda_2\gg\lambda_1$)
most state-changes occur after an effective local equilibration. There is just
one sequence of events that is able to avoid a clear local
equilibration. Namely that involving the forward jump ``$+$'' followed by the backward jump ``$-$'', which is the only sequence of transitions that allow the prefactor of the
second term to outweigh the first one. More precisely, using Eq.~(9) in the main text the corresponding waiting time density reads $\psi_{-|+}(t)=2.089\cdot 10^{-9}\e^{\lambda_1t}+6.738\cdot 10^{-3}\e^{\lambda_2t}$ with $6.738\cdot 10^{-3}\gg 2.089\cdot 10^{-9}$. Thus, after a short time $t\ll 0.15$, (i.e., much shorter than 15,\% of the mean waiting time) 
we  find a violation of local equilibration, whereas for $t\ge0.15$ a local equilibration is established, that is, $\psi_{-|+}(t)\approx2.089\cdot 10^{-9}\e^{\lambda_1t}$. Therefore, the violation of local detailed balance
leading to $\tilde\sigma_{1+2}/\sigma <1$ in Fig. 3a (for $\mu=20$) is caused almost entirely by the forward jumps ($+$) that are followed by a backward jump ($-$) which are shorter than 15\,\% of the total mean waiting time.

 Since backward transitions are extremely rare at $\mu=20$, we find
 that most transitions $>99.9\,\%$ occur upon a  local equilibration.
 
 Note that there are several other variants where local equilibration
 is manifested.  For example, using Eq. (3) in the main text one finds $P_{\rm even}\approx \frac{\kappa_+}{\kappa}\equiv P_{\rm even}^\text{eq}$
 whereas $P_{\rm odd}\not\approx1-\frac{\kappa_+}{\kappa}=\frac{\kappa_-}{\kappa}=P_{\rm odd}^\text{eq}$.
 The second line in Eq.~(5) 
 implies that the forward transitions are sampled from local equilibrium
 $W_+^{1+2}\approx P_{\rm even}^\text{eq} \omega_+$ but
 $W_-^{1+2}\not\approx P_{\rm odd}^\text{eq} \omega_-$. Thus, local
 equilibration sets in before each forward transition but \emph{not}
 before backward transitions.

 It is worth mentioning that if in addition to the time-scale
 separation \emph{all} individual transitions satisfy the mathematical
 stronger condition $\kappa_+,\kappa_-\gg \omega_+,\omega_-$ one
 restores local detailed balance in both directions (e.g., see
 Ref.~\cite{espo12}). In the following we discuss prominent examples
 from diffusion models which can \emph{never} satisfy this stronger condition.

%A separation of time-scales within the mesostate of Lumping $1+2$ demands $R\ll 1$. Let us now discuss the possible scenarios wether the follwing scenarios allow for $R\ll1$. We will discuss separately the scenarios (i) $\omega\sim \kappa$ (ii) $\omega\gg \kappa$ and (iii) $\omega \ll \kappa$.
%
%If (i) $\omega\sim \kappa$  we find in  Eq.~\eqref{eq:R} demands $q(1-q)\ll 1$, i.e., $q\approx0$ or $q\approx 1$. In addition we need to satisfy $(1-p)(1-q)+pq\ll 1$.
% If $q$
% 
% Since the cluster contain only two states, there will only one potentially quickly relaxing eigenvalue $\lambda_+$.
% Let us now quickly establish how the time scale 
% 

  \section{Kramers' theory and local detailed balance}
  In systems with continuous coordinates (for example, in any
  biophysical system) Markov-jump processes in the continuous-time
  limit (i.e., short lag-time limit) are well-known to not follow from
  a coarse-graining that is based on lumping \cite{sari10,schu11} [see
    also paragraph after Eq.~(2.47) in Ref.~\cite{moro95}].  In fact,
  we argue that local detailed balance may not exist if the
  coarse-graining is based solely on lumping. To see this we
  re-investigate below Kramers' original work  \cite{kram40} (see also
  Ref.~\cite{haen90} for a review), which pioneered the microscopic
  understanding of local detailed balance and,  more precisely, the
  constituents which are the rate velocities akin to transition rates.
  In stochastic thermodynamics Kramers' microscopic understanding is routinely used (directly or implicitly) to model stochastic pumps (cf. Arrhenius rates) \cite{astu03,raha08,espo15}, stochastic resonance \cite{qian00}, occasionally for kinetic proofreading \cite{ehre80,rao15}, and many others.

 \begin{figure*}
 \centering
 \includegraphics{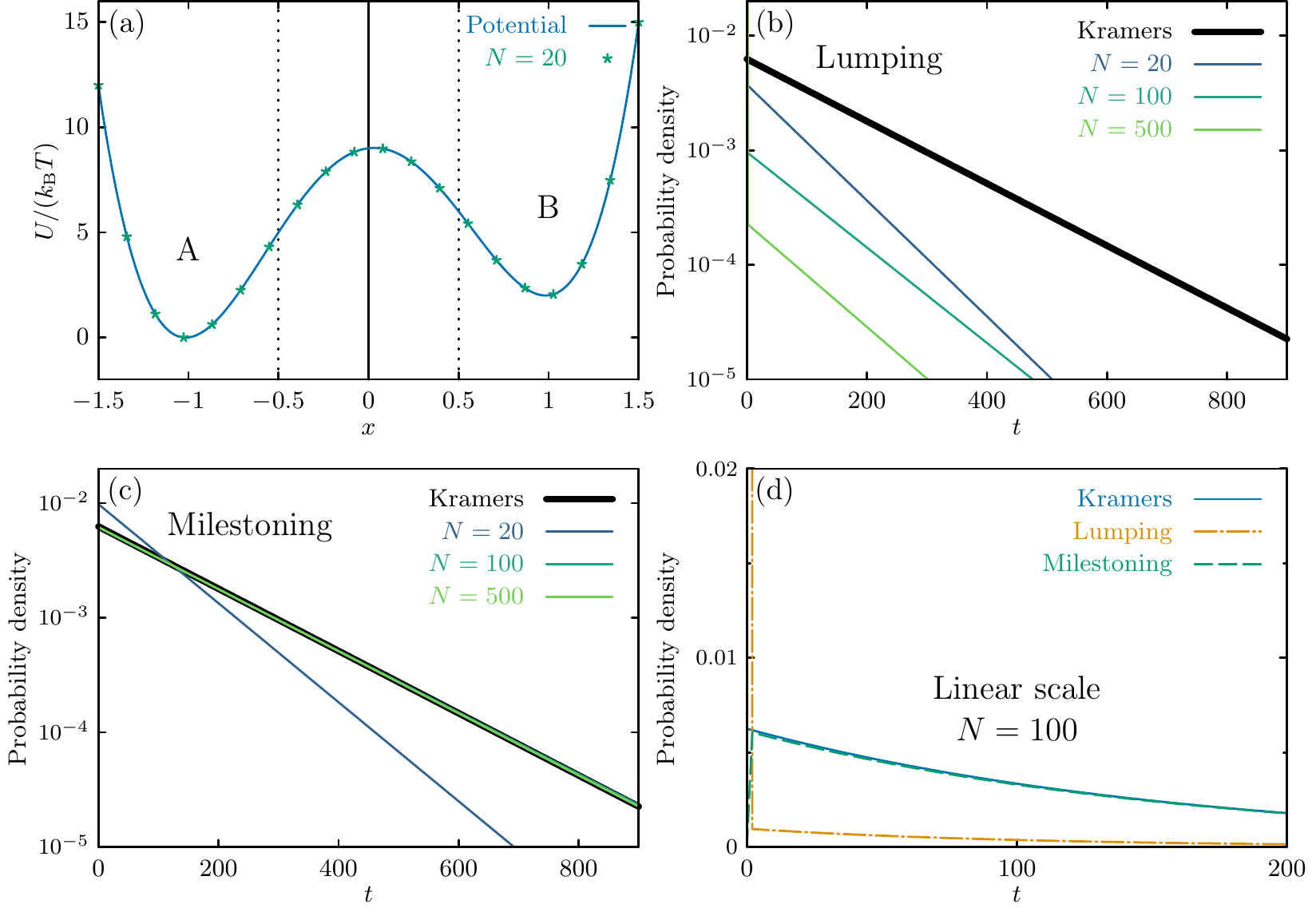}
 \caption{Testing Kramers' theory with a tilted double-well potential. (a)~Potential (solid blue line) and its discrete finite element (Markov state)
 representation with $N=20$ states. The vertical solid line  separates
 the potential into lumped states A and B. Milestoning is represented
 by the two vertical dotted lines within states A and B, where the
 region between the dotted lines can belong to either A or B
 (depending on which dotted line was crossed last).  (b)~Waiting time
 density in state B predicted by Kramers $\wp_{{\rm B}\to {\rm
     A}}=w_{\rm BA}\e^{-w_{\rm BA}t}$ using Eq.~\eqref{eq:wba_K}
 versus waiting time density to stay on the right half of the
 potential (vertical solid line in (a)) for $N=20,100,500$
 grid-points. (c)~First passage time density to pass the left dotted
 line at $x=-0.5$ for the first time if the particle starts right of
 the vertical dotted line at $x=0.5$ in (a), which represents the
 waiting time density in B based on Milestoning if the milestones are
 chosen by the vertical dotted lines in panel (b).  The thick line
 represents Kramers theory from (b). (d)~Direct comparison of Kramers
 theory with Lumping and Milestoning with a finite element method and
 $N=100$ states.  In the limit $N\to\infty$ Lumping leads to delta
 distribution of the probability density with the probability weight
 moving to $t=0$.}
 \label{fig:Kramers_test}	
 \end{figure*}

In the following we rationalize why Kramers' theory
\cite{kram40,gard04} is incompatible with lumping in continuous time
processes. To illustrate this we consider a particle at position $x$
in a tilted double-well potential
\begin{equation}
	\frac{U(x)}{k_{\rm B}T}=8*(x+1)^2+(x-1)^2+x,
	\label{eq:U}
\end{equation}
where the last term introduces the tilt. The potential is illustrated
in Fig.~\ref{fig:Kramers_test}a, where the vertical solid line
separates the two lumped meso-states which include the minima A and B
(the vertical dotted lines will be used for Milestoning below).  The
tilt leads to a free energy difference between right state B and left
state A to be given by $\approx 2\,k_{\rm B}T$.

To use Kramers' theory we determine the location of two minima of the potential \eqref{eq:U} at
$x_{\rm A}\approx -1.01527$ and $x_{\rm B}\approx0.983993$
alongside the local maximum at $x_{\rm max}=0.0312806$. For an overdamped particle with diffusion constant set to unity $D=1$ and $k_{\rm B}T\equiv 1$ Kramers' work
predicts the rates to be approximately given by \cite{kram40,gard04}
\begin{align}
	w_{{\rm A}\to {\rm B}}&=\frac{\sqrt{U''(x_{\rm A})U''(x_{\rm max})}}{2\pi}\e^{U(x_{\rm A})-U(x_{\rm max})}
	\nonumber\\
	&
	\approx 8.8689\times10^{-4}
	\label{eq:wab_K}
\end{align}
and
\begin{align}
	w_{{\rm B}\to {\rm A}}&=\frac{\sqrt{U''(x_{\rm B})U''(x_{\rm max})}}{2\pi}\e^{U(x_{\rm B})-U(x_{\rm max})}\nonumber\\
	&\approx 6.2510\times10^{-3},
	\label{eq:wba_K}
\end{align}
where $U''(x)\equiv\partial_x^2U(x)$ is the second derivative that
relates to the curvature of the potential. In the following we test
Kramers' result first against lumping and then against Milestoning.

\subsection{Lumping is inconsistent with Kramers' theory}
Let us first test Lumping against Kramers'  rates. To test this
numerically we use a thermodynamically consistent  finite element
method from \cite{holu19} which we also used in \cite{lapo20} to
derive local time distribution for diffusion processes.  
To this end we discretize our potential in $N$ steps. The discretization $i=1,\ldots ,N$ for $N=20$ is shown in Fig.~\ref{fig:Kramers_test}a (see green symbols).
For convenience we confine the grid between equidistant points
$x_1=-1.5$ and $x_N=1.5$ with a grid spacing $dx=x_{i+1}-x_{i}=(x_N-x_1)/(N-1)=3/(N-1)$.
The finite-element method (FEM) is chosen to represent a Markov-jump process between the states $i$ and $j=i\pm 1$ for any $1\le i,i\pm1\le N$ to be given by \cite{holu19} (see also \cite{lapo20})
\begin{equation} 
	w_{i\to i\pm 1}^{\rm FEM}=\frac{D}{(dx)^2}\exp\left[\frac{U(x_i)-U(x_{i\pm1})}{2k_{\rm B}T}\right],
	\label{eq:FEM}
\end{equation}
where we set $D=k_{\rm B}T=1$. 

To evaluate the waiting time in B for a given grid with $N$ states
based on lumping we measure the waiting time of the process until the
right half of the potential (right of vertical dotted line in
Fig.~\ref{fig:Kramers_test}a) is left for the first time after
entering said region.  Using the rates from Eq.~\eqref{eq:FEM} leads to the waiting time density in Fig.~\ref{fig:Kramers_test}b for $N=20,100,500$.
The following two observations can be made.

First, the waiting time distribution becomes shifted towards shorter
times as $N$ increases, which we mathematically expect,  since any
overdamped diffusion process (i.e., $N\to\infty$) formally
``wriggles'' infinitely often across any position it visits, which in
turn leads to vanishingly short waiting times in the limit
$N\to\infty$.  Note that even underdamped systems can wriggle several
times across a barrier, which is typically accounted for  by the
success probability to ``actually'' cross the barrier also called the
``transmission coefficient'' (e.g., see Ref.~\cite{elbe20}).   It is
worth mentioning that we believe that ``Dynamical coring''  \cite{jain14,nage19} tackles this problem by effectively removing all the fast fast re-crossings, which means the waiting time density then approximately selects the rate of the long time tail of the lumped process. 

Second, excluding the fast re-crossings and focusing on the long-time
limit we find a non-normalized exponential decay that is faster than
Kramers' result, which is due to the fact that any lumped state
``senses'' only one half of the barrier's curvature which enters
Eqs.~\eqref{eq:wab_K} and \eqref{eq:wba_K}.  The full curvature requires a full crossing of the barrier, which we discuss below in more detail whilst discuss Milestoning. Note that Fig.~\ref{fig:Kramers_test}b
shows the probability density on a semi-log-scale, which renders exponential decays to straight lines. See Fig.~\ref{fig:Kramers_test}d for the linear-scale plot with $N=100$, which also includes Milestoning that is discussed in the following subsection.

These aforementioned two aspects the (re-crossing \emph{and} half of
barrier sampling) are the key reasons why Kramers' theory is
fundamentally incompatible with lumping. 
Note that we are not the first to show this incompatibility, since it
follows from several previous findings, which can in particular be
drawn from the paragraph following Eq.~(2.47) in Ref.~\cite{moro95}
and is also mentioned in Refs.~\cite{sari10,schu11} to name but a few.
In contrast to our work pursued the main text, this conflict arises
merely from projecting a \emph{continuous} space dynamics into
\emph{discrete} one, which we also pointed out in our recent work
\cite{hart21}.

  \subsection{Milestoning in Kramers' theory}
Let us now study Kramers' rate theory in the context of Milestoning.
To this end we locate the two milestones at $x=-0.5$ (A) and $x=0.5$
(B) (see vertical dotted lines in Fig.~\ref{fig:Kramers_test}a). The
coarse-graining that is based on Milestoning simply maps the full
dynamics into the last crossed milestone. Thus, for any $x\le -0.5$
the coarse-grained state is ``A'' and for any $x\ge 0.5$ the state is
mapped to ``B'', whereas for any $-0.5\le x\le 0.5$ the process is
mapped to either ``A'' or ``B'', depending on which milestone was
passed last.  Note that we start to record the coarse-grained state as
soon as the first milestone is crossed. One waiting time interval in
state B corresponds to the time between the first entrance into B by
crossing $x=0.5$ until the other milestone A at $x=-0.5$ is crossed
for the first time.  

To measure the waiting time we use the finite element method from
Eq.~\eqref{eq:FEM}. To deduce the waiting time density in state B we
use the grid state of the right dotted line ($x=0.5$) as the starting
condition and the state left of the left dotted line ($x=-0.5$) as the
target or absorbing condition.  The resulting waiting-time density is shown in Fig.~\ref{fig:Kramers_test}c. 
We find for $N\gtrsim20$ that the waiting time density to a good approximation approaches the waiting time density predicted by Kramers approximation, i.e., Arrhenius law. We show the waiting time density for $N=100$
in Fig.~\ref{fig:Kramers_test}d on a linear scale (see Milestoning and Kramers). 
For completeness we show the mean values of the waiting time including
the reverse transition ${\rm A}\to {\rm B}$ in Tab.~\ref{tab:MWT}. One
can see that the mean waiting time determined by Milestoning
approximates the Kramers result reasonably well.

\begin{table}
  \caption{Mean waiting time in A and B. Kramers approximation  follows from the inverse  of the rates $w_{{\rm A}\to{\rm  B}}$
and $w_{{\rm B}\to{\rm  A}}$ from \eqref{eq:wab_K} and \eqref{eq:wba_K}, respectively, which are used as a reference. The relative deviation from Kramers is $\epsilon=|w^{-1}-\avg{t}|/w^{-1}$, where $w^{-1}=w_{{\rm A}\to{\rm  B}}^{-1}, w_{{\rm B}\to{\rm  A}}^{-1}$ and $\avg{t}=\avg{t}_{\rm A},\avg{t}_{\rm B}$. The mean waiting time evaluated for the two variants of coarse-graining (Lumping and Milestoning)  and for serveral  discretisations $N=20,100,500$. }
\begin{tabular}{c|c|c}
	&\multicolumn{2}{c}{Waiting time in state A}\\\cline{2-3}    
	&Mean value $\avg{t}_{\rm A}$& rel. error $\epsilon$\\\hline
Kramers $w_{{\rm A}\to\rm B}^{-1}$&\bfseries 1127.5&\bfseries 0\,\%\\
$N=20$ (Lumping)&189.04&83.2\,\%\\
$N=100$ (Lumping)&74.556&93.4\,\%\\
$N=500$ (Lumping)&15.346&98.6\,\%\\
$N=20$ (Milestoning)&678.07&39.9\,\%\\
$N=100$ (Milestoning)&1148.0&1.81\,\%\\
$N=500$ (Milestoning)&1172.4&3.98\,\%\\
\hline\hline
	&\multicolumn{2}{c}{Waiting time in state B}\\\cline{2-3}
	&Mean value $\avg{t}_{\rm B}$& rel. error $\epsilon$\\\hline
Kramers $w_{{\rm B}\to\rm A}^{-1}$&\bfseries 159.97&\bfseries 0\,\%\\
$N=20$ (Lumping)&27.604&82.7\,\%\\
$N=100$ (Lumping)&74.556&93.4,\%\\
$N=500$ (Lumping)&2.1256&98.7,\%\\
$N=20$ (Milestoning)&98.977& 38.1\%\\
$N=100$ (Milestoning)&159.82&0.098,\%\\
$N=500$ (Milestoning)&158.85& 0.70\,\%\\
\end{tabular}
\label{tab:MWT}
\end{table}

\subsection{Why is Arrhenius law consistent with Milestoning?}
To see this we look at its derivation \cite{kram40,gard04} based on a
first passage problem.  In Milestoning the pair of vertical dotted
lines correspond to the starting and target points, respectively. In the case of an escape from A the dotted line at $x=-0.5$ would be the initial condition and $x=0.5$ would be the absorbing target, whereas for an escape from B the lines exchange roles (target and starting point). 

Eq.~\eqref{eq:wab_K} relates to Eq.~(5.2.174) in Ref.~\cite{gard04}, which  uses the starting point to be at the left minima and the target position (absorbing point)
the well on the right half of the barrier (e.g., the right minima).
The exact location of the target point does matter much (see Fig.~5.3c
in Ref.~\cite{gard04})  as soon as the target is well to the right
from the barrier (local maximum $x\approx 0$). In other words, the
location of the right dotted line will not substantially affect the
Milestoning process as  long as its location is right of the barrier
\emph{and} a few thermal energy units below its maximum value.  This
guarantees that the particle will revisit the Milestone several times
before hitting the other Milestone, which is the key ingredient for
the process to become approximately Markovian \cite{schu11} and allows
the particle to locally relax to the equilibrium Boltzmann
distribution within the minimum B. 

Due to the equivalence between the first passage problem proposed by
Kramers and the Milestoning process, it is obvious that Milestoning is
consistent with Kramers theory.

\subsection{Closing remarks}
A few important aspects should be pointed out.

First, despite the substantial quantitative difference in the waiting
time density between the variants of coarse-graining (Lumping and
Milestoning) there are a few strong similarities that we need to point
out. If we evaluate both coarse-graining procedures on a single trajectory we will find that most of the time (that is more than 99\,\% of the entire time) we expect the microstate to be mapped to the same mesostate (A, B). The reason is that Milestoning and Lumping can only lead to a different coarse-grained process if the particle is located between $-0.5\le x\le 0.5$, which rarely happens.

First, due to the high energy within said region the particle will only spend a very short time within that region. Second, whenever the particle is on the left half $-0.5\le x\le 0$ it will have much more likely passed $x=-0.5$ for the last time and not  $x=0.5$, which further decreases the likeliness at any time $t$ that both coarse-graining procedures (Lumping or Milestoning) map the microstate into a different Mesostate. Thus, the difference between Lumping and Milestoning can only be detected between fast transition-path  events.

Therefore, if an experiment has a time resolution in the form of
discrete time intervals that cannot resolve transition-path events, we
do not expect the experiment to detect any recrossing events and
thus to actually detect the violation of local detailed balance of a
Lumped process.

Note that a careful reader might use Tab.~\ref{tab:MWT} and evaluate the mean waiting time according to Lumping in A and B, respectively, and deduce from its inverse that Eq.~(1) in the main text is approximately satisfied
\begin{equation}
	k_{\rm B}T\ln \frac{\avg{t}_{\rm A}}{\avg{t}_{\rm B}}\approx 2k_{\rm B}T
\end{equation}
for any grid size $N$ tacitly assuming the rates  with Lumping are taken as the inverse of the mean $w_{{\rm A}\to {\rm B}}^{\rm Lumping}=1/\avg{t}_{\rm A}$ and $w_{{\rm B}\to {\rm A}}^{\rm Lumping}=1/\avg{t}_{\rm B}$.
We expect this ``coincidence'' to vanish for genuinely out-of-equilibrium systems
\begin{equation}
	\ln\frac{w_{{\rm B}\to {\rm A}}}{w_{{\rm A}\to {\rm B}}}
	\neq\ln \frac{P_{\rm A}^{\rm ss}}{P_{\rm B}^{\rm ss}}.
\end{equation}

We were recently able to show that Milestoning  with a time-scale separation arbitrarily far from equilibrium is consistent with Kramers' theory \emph{and} satisfies local detailed balance (see Sec. III.C ``The peculiar limit of local detailed balance'' in \cite{hart21}). Consistent with Ref.~\cite{hart21}, our results show, for the first time, that milestoning may be crucial to restore local detailed balance for for coarse-graining of Markov jump processes along individual trajectories.

\section{Lumping is a subset of Milestoning}
%Let us briefly explain why Lumping is a subset of Milestoning. In fact any Lumped process can be obtained in as the limit of Milestoning. We will exemplary discuss a few examples.
There is one freedom that Milestoning allows which is not allowed in
Lumping. Lumping asserts to any microstate a single coarse-grained
mesostate, whereas Milestoning allows for an ``neutral''  region of
microstates that are not uniquely attributed to a specific single
coarse-grained mesostate \cite{schu11}. Note  within such a
``neutral''  region Milestoning determines the coarse-grained state by
the last visited non-neutral region.  In this sense Lumping is the
limit of Milestoning, where the ``neutral region'' is the empty set,
which completes the proof that Lumping is a subset of Milestoning.

Note that the aforementioned neutral region in the main text
represents the space between the two ellipses in Fig.~1b, or the space between the two lines in Fig.~1c.
The gray states in Fig. 2 in the main text represent such  ``neutral''
states, and similarly, the space between vertical dotted lines in
Fig.~\ref{fig:Kramers_test}a (i.e., $-0.5\le x\le 0.5$).

%In the colorcode of the trajectory in Fig.~1a displays a process based on Lumping while Fig.~1b shows an example for Milestoning. Lumping does not allow for regions that do not define the state precisely.
%Howecer, if we use the Milesones (here ellipses) from Fig.~1b and ``blow them up'' and reshape  them to ``touch'' the separation line as depicted in Fig.~1a in the main text from both sides, we will effectively will effectively formulate the lumped process as a Milestoning one. This completes the proof for this example that we can find peculiarly chosen milestones that emulate lumping. The reverse is not always possible.

%Kramer's microspopic underpinning is routinily implicitly used in numerous contexts to model 

%  It proofs convendient to first indentfy $\lambda_++\lambda_-=\omega+\kappa_-$ and then 
% 
% to translate the former condition into $\lambda_+\lambda_-/(\omega)\ll$ 
%% \end{acknowledgments}

  \input{bib_main.bbl}
%\bibliography{/Users/hartich/Dropbox/ausgewertete_Litaratur/Bib/Biballes}
 \end{document}